# THE BRIGHTEST STARS IN GALAXIES ARE NOT GOOD DISTANCE INDICATORS

Richard Rozanski[1], Michael Rowan-Robinson[1]

[1] *Astronomy Unit, School of Mathematical Sciences, Queen Mary & Westfield College, Mile End Road, London, E1 4NS, UK.*

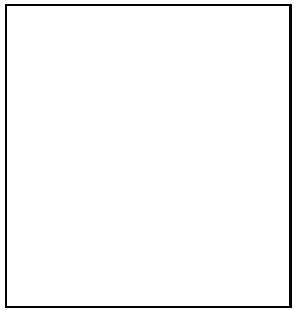

**Abstract**

Use of the brightest stars in galaxies as a distance indicator and claims that the method supports the 'short' distance scale are examined. Data from several different observational programmes are brought together for the first time with a procedure for the careful accounting of errors. The true uncertainties of the method are found to be much larger than claimed by its advocates. The method is incapable of distinguishing between the 'long' and 'short' distance scales.

## 1  Introduction

The use of the brightest stars in galaxies as distance indicators is based on the assumption that their intrinsic luminosity is a predictable quantity. In practice the method is applied to the following classes of stars in spiral and irregular galaxies: blue supergiants of spectral classes O, B, A with $B - V \lesssim 0.4$ and red supergiants of spectral classes K5–M5 with $B - V \gtrsim 2.0$. The colour constraints were first proposed by Sandage & Tammann [14] to help avoid confusion with foreground stars.

An enormous amount of observational data now exists on the brightest stars in many nearby galaxies. Unfortunately this data has never been brought together in order to examine the method's reliability and properly assess its uncertainties. Its principle advocates, Allan Sandage and Roberta Humphreys, each tend to use only their own observations in calibrating the method. Their respective calibrations can be found in Humphreys [3, 4] and Sandage & Carlson [13].

# 2 The nature of the calibration

This paper reports on an effort to bring together as much of the observational data as possible and examine whether the method is as accurate as claimed. In order to avoid excessive distortion of the calibration if the wrong brightest candidate star is picked, the calibration uses the average of the magnitudes of the three brightest stars: $m_\star(3)$. Using galaxies of known distance a plot of the following form can be produced to examine any dependence on parent galaxy luminosity: for blue stars $M_b(3)\,v.\,M_{gal}$; for red stars $M_v(3)\,v.\,M_{gal}$ and $M_k(3)\,v.\,M_{gal}$. If a linear relationship in a given band is found then an alternative calibration can be obtained correlating $M_{gal}\,v.\,(m_\star(3) - B_{gal})$. The advantage of this form of calibration is that it avoids the effects of distance degeneracy – see point 3 below. A number of points need to be made about the nature of the calibration:

1. In each case where a non-zero slope exists the equation needs to be transformed as follows in order to obtain the distance modulus of a programme galaxy:

$$\mu_\circ = \frac{m_\star(3)}{(1-\mathbf{a})} - \frac{\mathbf{a}B_{gal}}{(1-\mathbf{a})} - \frac{\mathbf{b}}{(1-\mathbf{a})} \qquad (1)$$

   where $\mathbf{a}$ is the slope and $\mathbf{b}$ is the intercept of the calibrating plot respectively.

2. If the r.m.s. scatter about the mean regression is given by $\sigma$ then a non-zero slope increases the error in $\mu_\circ$ due to the effects of distance degeneracy. This is illustrated in Figure 1. The size of $\delta\mu_\circ$ is given by:

$$\delta\mu_\circ = \frac{\sigma}{(1-\mathbf{a})} \qquad (2)$$

   This is the minimum error since the photometric errors in $m_\star(3)$ and $B_{gal}$ also propagate into $\mu_\circ$ as can be seen from equation 1.

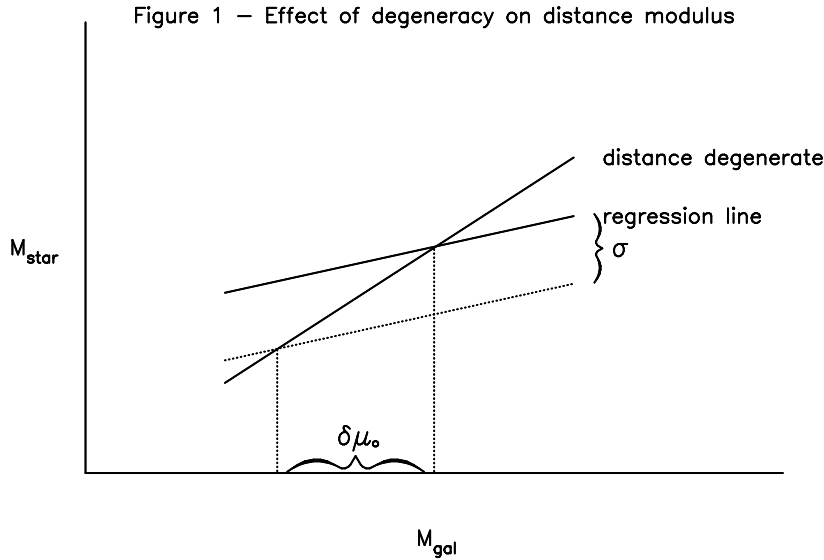

**Figure 1** – Relationship between the regression error $\sigma$ on the slope relating $M_\star\,v.\,M_{gal}$ and the resulting error in the estimated distance modulus. If $\mathbf{a}$ is the gradient of the regression line then $\delta\mu_\circ = \sigma/(1-\mathbf{a})$.

3. Each form of the calibration requires careful definition. The blue stars must be the brightest in the $B$ band. If the $V$ band is to be used then it must be made clear if the stars being used are the 3 brightest in $V$ or if the $V$ mags of the 3 brightest stars in $B$ are being used. For the red stars this is even more important. The 2 bands in which most observations have been done are the $V$ and $K$ bands. For the $V$ band the stars must be the brightest in $V$. For the $K$ band it is necessary to be clear whether the stars come from list of stars brightest in $V$ or whether *only* $K$ band mags are taken into account. In most of the observational work done so far the former approach has been adopted. This means that any $K$ band photometry must be carried out on pre-existing lists of stars brightest in $V$ or in conjunction with $V$ band observations. Most importantly it must be understood that these are the only observational bands and classes of stars for which calibrations exist. Observations in $I$ or $R$ bands *cannot* be used in conjunction with the $B$, $V$ or $K$ band calibrations because the brightest stars at $I$ or $R$ are not necessarily the brightest at $V$, $K$ or $B$.

# 3 Recent results using the method

A number of papers have been published recently which use the method on 2 Virgo group galaxies and on IC 4182. The significance of these papers is that the results are used to constrain the value of $H_\circ$. In all 3 cases the results appear to support the 'short' distance scale – see Jacoby et al. [5] and van den Bergh [18] for recent reviews supporting this position.

## 3.1 NGC 4523

Shanks et al. [17] report $R$ band observations of yellow supergiants in NGC 4523. They obtain $\mu_{\rm AR} = 30.6 \pm 0.3$. As explained above there is no calibration for any class of stars in the $R$ band nor for yellow supergiants. Shanks et al. are therefore forced to use only *one* galaxy – the LMC – to calibrate their observations. As Section 5 shows, the inherent scatter in the method is much higher than 0.3 mag for classes of stars and observational bands for which a proper calibration exists. Therefore this result is an inappropriate use of the method with a grossly underestimated uncertainty.

## 3.2 NGC 4571

Pierce et al. [6] report $R$ band observations of the brightest blue and red supergiants in NGC 4571. They obtain $\mu = 30.9 \pm 0.2$. This result is also based on an unjustified extrapolation of the blue and red star calibrations into the $R$ band using assumed colours.

## 3.3 IC 4182

Pierce et al. [7] report $I_{kc}$ and $K$ band observations of red supergiants with $\mu = 27.0 \pm 0.2$. This is discussed further in Section 5.1 below.

# 4 Procedure adopted to assess the method

In an attempt to assess the reliability of the method we have adopted the following procedure:

1. Bring together data from as many observational programmes and reduce them to a common scale with careful accounting of errors.

2. For each paper select the brightest 3 candidates and obtain $m_\star(3)$ and $\sigma$ where:

$$\sigma = [\sigma_{phot}^2 + \sigma_{spread}^2]^{\frac{1}{2}} \quad (3)$$

$$\sigma_{phot}^2 = \frac{\Sigma_{i=1}^3 \Delta m_i}{3} \quad (4)$$

$$\sigma_{spread}^2 = \frac{\Sigma_{i=1}^3 (m_i - m_\star(3))^2}{2} \quad (5)$$

This is the first time that any error has been attached to the quantity $m_\star(3)$. Previously all the advocates of the method have ignored this.

3. For each galaxy combine the estimates of $m_\star(3)$ weighted using $1/\sigma^2$ to obtain $\overline{m_\star(3)}$, $\sigma_{int}$ and $\sigma_{ext}$ for each galaxy.

4. Obtain the least squares regression using $\sigma_{int}$ as weighting. The final $\sigma$ error is the r.m.s deviation of the points about the regression line.

5. Corrections for *internal* extinction have proved controversial. Sandage [12] argues that a correction is unnecessary since it will be low and/or it should be the same as for the Population I objects – ie. Cepheids – used to find the calibrating galaxies' distances. Humphreys [2] disagrees, arguing that ignoring internal extinction distorts the calibration since it will not necessarily be low or uniform because the OB associations which produce these stars tend to be in cloudy regions. Where individual measurements via spectroscopy and/or multicolour photometry are available they have been applied here.

6. Where no internal extinction has been measured then a correction has been made for *foreground* extinction. This is taken from the IRAS 100 micron all-sky map – see Rowan-Robinson et al. [10] for details. Hereafter these are indicated using either $A_{b\,I(100)}$ or $A_{v\,I(100)}$. These corrections have also been applied to galaxy apparent magnitudes which are taken from the Third Reference Catalogue of Bright Galaxies [1]. The errors on the foreground extinction measurements are as follows: $\sigma(A_b) = 0.09$, $\sigma(A_v) = 0.07$, $\sigma(A_k) = 0.01$.

| 1 | 2 | 3 | 4 | 5 | 6 | 7 | 8 | 9 | 10 | 11 |
|---|---|---|---|---|---|---|---|---|---|---|
| Name | $\mu_\circ$ | $M^\circ_{bT}$ | $\sigma(M^\circ_{bT})$ | $A_{b\,I(100)}$ | $M_v(3)$ | $\sigma_{int}$ | $M_k(3)$ | $\sigma(K)$ | $M_b(3)$ | $\sigma_{int}$ |
| LMC | 18.64 | -18.04 | 0.14 | 0.24 | -7.93 | 0.21 | -11.94 | 0.22 | -9.42 | 0.40 |
| SMC | 19.09 | -16.74 | 0.16 | 0.11 | -7.94 | 0.27 | -11.39 | 0.19 | -9.14 | 0.38 |
| I1613 | 24.21 | -14.50 | 0.16 | 0.12 | -7.58 | 0.41 | -10.81 | 1.13 | -7.56 | 0.26 |
| M31 | 24.25 | -20.92 | 0.14 | 0.36 | -8.23 | 0.43 | -11.37 | 0.29 | -8.10 | 0.30 |
| M33 | 24.43 | -18.77 | 0.19 | 0.28 | -8.52 | 0.19 | -11.62 | 0.38 | -9.80 | 0.41 |
| N6822 | 23.55 | -15.15 | 0.20 | 0.92 | -8.04 | 0.42 | -11.28 | 0.74 | -8.64 | 0.45 |
| W-L-M | 24.72 | -14.25 | 0.16 | 0.21 | -6.62 | 0.27 | – | – | -7.19 | 0.26 |
| SexA | 26.01 | -14.40 | 0.23 | 0.19 | -7.81 | 0.31 | -11.27 | 0.66 | -7.38 | 0.30 |
| SexB | 25.92 | -14.59 | 0.26 | 0.28 | -7.5 | 0.24 | – | – | -6.75 | 0.35 |
| N300 | 26.09 | -17.67 | 0.14 | 0.09 | -7.8 | 0.73 | -11.11 | 0.24 | – | – |
| N3109 | 26.0 | -16.71 | 0.15 | 0.37 | -7.57 | 0.42 | -11.67 | 0.52 | -7.57 | 0.34 |
| N2403 | 27.43 | -19.07 | 0.23 | 0.24 | -7.84 | 0.41 | -10.93 | 0.27 | -9.23 | 0.37 |
| M81 | 27.43 | -20.35 | 0.22 | 0.47 | -8.43 | 0.46 | -11.49 | 0.42 | -9.82 | 0.50 |
| N2366 | 27.43 | -16.53 | 0.24 | 0.23 | -6.96 | 0.26 | – | – | -8.18 | 0.37 |
| I2574 | 27.43 | -17.23 | 0.29 | 0.20 | -7.33 | 0.26 | – | – | -7.92 | 0.32 |
| N4236 | 27.43 | -17.96 | 0.28 | 0.12 | -7.71 | 0.24 | – | – | -8.39 | 0.30 |
| N1560 | 27.43 | -17.03 | 0.26 | 0.66 | -7.31 | 0.22 | – | – | -8.15 | 0.28 |
| N2976 | 27.43 | -17.40 | 0.25 | 0.37 | -6.81 | 0.32 | – | – | -8.32 | 0.42 |
| DDO165 | 27.43 | -14.74 | 0.30 | 0.13 | -6.60 | 0.23 | – | – | -7.24 | 0.47 |
| HoI | 27.43 | -14.59 | 0.30 | 0.23 | – | – | – | – | -7.83 | 0.49 |
| HoII | 27.43 | -16.61 | 0.27 | 0.17 | -7.33 | 0.24 | – | – | -7.93 | 0.25 |
| HoIX | 27.43 | -13.57 | 0.37 | 0.38 | -7.24 | 0.47 | – | – | -6.91 | 0.67 |
| M101 | 29.24 | -21.20 | 0.24 | 0.16 | -8.98 | 0.38 | -13.17 | 0.35 | -10.4 | 0.56 |
| N5474 | 29.24 | -18.05 | 0.27 | 0.11 | – | – | – | – | -8.8 | – |
| N5585 | 29.24 | -18.30 | 0.26 | 0.11 | – | – | – | – | -8.4 | – |
| I4182 | | | 0.72 | 0.09 | | | | | | |

Table 1 – Summary of Data for Local Group & other nearby galaxies

**Column entries for Table 1:** 1). Galaxy name. 2). Distance modulus. These are all taken from Rowan-Robinson [8, 9]. All galaxies in the NGC 2403/M81 and M101 groups are assigned mean group distances. 3). $M^\circ_{bT}$ is the absolute corrected galaxy magnitude used in the calibration. 4). $\sigma(M^\circ_{bT}) = [\sigma^2(B_{gal})+\sigma^2(A_{b\,I(100)})+\sigma^2(\mu_\circ)]^{\frac{1}{2}}$. Used to weight regression fits. Errors taken from [1] and [9]. 5). $A_{b\,I(100)}$ is the foreground extinction value taken from 100 micron flux all-sky maps. 6). $M_v(3)$ is the corrected absolute $V$ band average for red supergiants. 7). $\sigma_{int}$ is the internal error on $M_v(3)$ used to weight the regression fits. 8). $M_k(3)$ is the corrected absolute $K$ band average for red supergiants. 9). $\sigma(K)$ is the error on $M_k(3)$ used to weight the regression fits. 10). $M_b(3)$ is the corrected absolute $B$ band average for blue supergiants. 11). $\sigma_{int}$ is the internal error on $M_b(3)$ used to weight the regression fits.

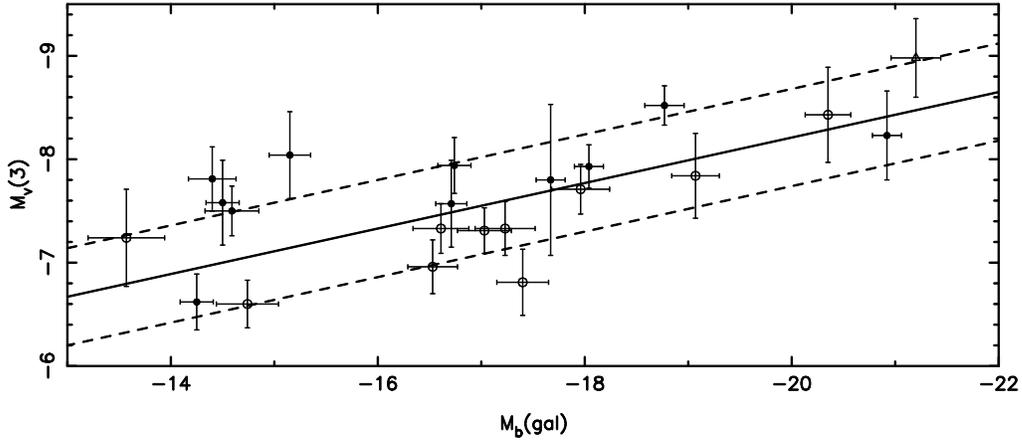

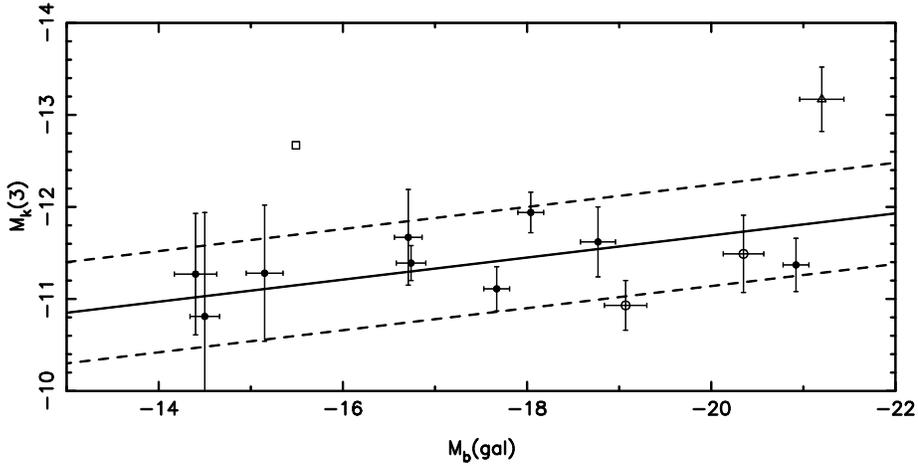

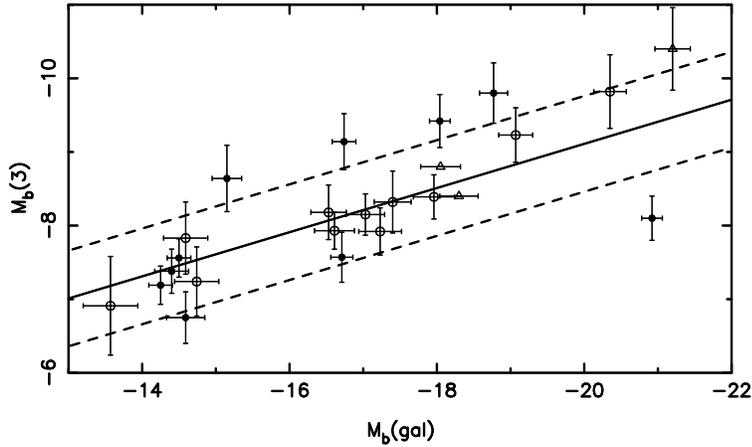

**Figure 2** – $V, K$ and $B$ band luminosity calibrations for the brightest supergiants in each sample galaxy. The filled circles are the Local Group, NGC 300, NGC 3109, Sextans A and Sextans B. The open circles are NGC 2403/M81 group galaxies. The open triangles are M101 group galaxies. The 2 open triangles with horizontal error bars only are the 2 M101 group galaxies NGC 5474 and NGC 5585 which are not used in the calibration. The open square in Figure 2b gives the locus of IC 4182 if the Pierce et al. $K$ band data is combined the the Cepheid distance of Sandage et al. The solid line is the regression line whose coefficients are given in Table 2. The dashed line is the $1\sigma$ regression error.

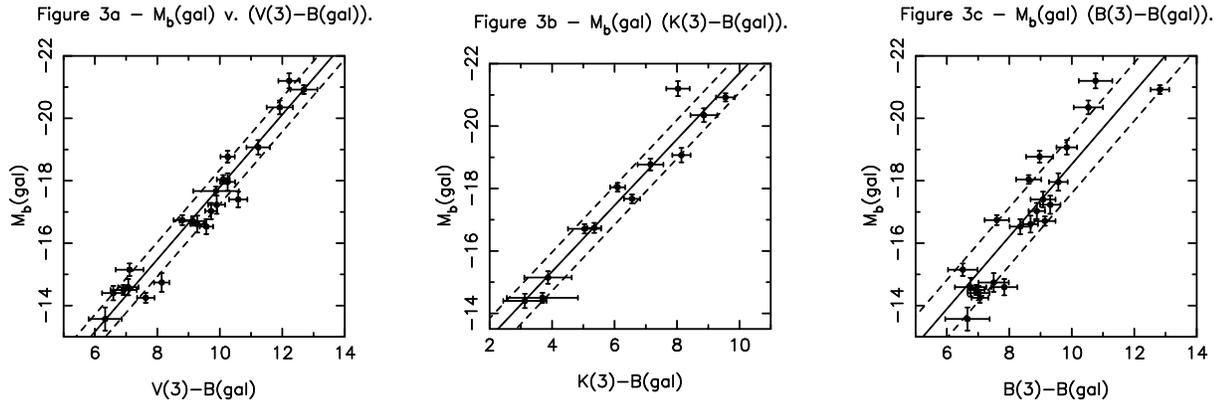

**Figure 3** – Calibration of the apparent magnitude difference: $(V(3) - B_{gal})$ for red supergiants; $(K(3) - B_{gal})$ for red supergiants; $(B(3) - B_{gal})$ for blue supergiants. All galaxies have the same symbol. The solid and dashed lines are as for Figure 2.

| Table 2 – Coefficients for least squares fits | | | | | | | | |
|---|---|---|---|---|---|---|---|---|
| 1 | 2 | 3 | 4 | 5 | 6 | 7 | 8 | 9 |
| Fig/Reg. no. | N | **a** | **b** | **c** | **d** | **e** | $\sigma$ | $\delta\mu_o$ |
| 2a | 22 | $0.22 \pm 0.03$ | $-3.81 \pm 0.57$ | 1.28 | -0.28 | 4.91 | 0.47 | 0.60 |
| 2b | 12 | $0.12 \pm 0.05$ | $-9.29 \pm 1.01$ | 1.14 | -0.14 | 10.56 | 0.55 | 0.62 |
| 2c | 22 | $0.30 \pm 0.04$ | $-3.11 \pm 0.62$ | 1.43 | -0.43 | 4.44 | 0.65 | 0.93 |
| 3a | 22 | $-1.16 \pm 0.02$ | $-6.20 \pm 0.23$ | – | – | – | 0.55 | 0.55 |
| 3b | 12 | $-1.06 \pm 0.02$ | $-11.10 \pm 0.17$ | – | – | – | 0.62 | 0.62 |
| 3c | 22 | $-1.16 \pm 0.02$ | $-6.93 \pm 0.22$ | – | – | – | 0.93 | 0.93 |

**Column entries in Table 2:** 1). Figure/regression number. 2). N is the number of data points used in the regression. 3). **a** is the gradient of regression line and standard error. 4). **b** is the intercept of regression line and standard error. 5, 6 & 7). **c, d, e** are the coefficients of $m_\star(3)$ and $B_{gal}$ and zero-point in equation 1 – applies only to regressions 2a, 2b & 2c. 8). $\sigma$ is the r.m.s deviation of the data points about the regression line. 9). $\delta\mu_o$ is the minimum error on the distance modulus obtained using the regression as explained above.

## 5 Results and discussion

The full list of papers examined can be found in Rozanski & Rowan-Robinson [11]. Table 1 shows the results of applying the procedure to each galaxy in the calibrating sample. In addition parameters are given for IC 4182 observed by Pierce et al.

The calibrating plots are shown in Figures 2 & 3. Table 2 shows the coefficients obtained for the various calibrations using least squares fits.

A number of points are apparent:

1. For the blue stars the dependence of $M_b(3)$ on $M_{gal}$ is as found by both Humphreys and Sandage.

2. For the red stars both the $V$ and $K$ band calibrations show a non-zero slope.

3. The scatter for the red stars is higher than reported by either Humphreys or Sandage. The minimum distance modulus error using this method is 0.55 mag using $(M_v(3) - B_{gal})\,v.\,M_{gal}$.

4. The distance modulus error listed in column 9 is a *minimum* error which does not include photometric errors in the observed quantities $m_\star(3)$ and $B_{gal}$.

## 5.1 IC 4182

The $K$ band observations of Pierce et al. can now be used in conjunction with the $K$ band calibrations obtained above. Table 3 gives the results. Note that the error in $K_\circ(3)$ has been obtained using the method described in Section 4.

| Table 3 – K band distance estimates to IC 4182 | | | |
|---|---|---|---|
| | | $\mu_\circ$ | |
| $B_T^\circ$ | $K_\circ(3)$ | Reg.2b | Reg.3b |
| $12.91 \pm 0.72$ | $15.73 \pm 0.11$ | $26.68 \pm 0.62$ | $27.00 \pm 0.95$ |

These results can be compared with results obtained using the $V$ and $B$ band photometry presented in Sandage & Tammann [15]. This is given in Table 4. Note that they give no error estimates for $V(3)$ and $B(3)$.

| Table 4 – V and B band distance estimates to IC 4182 | | | | | |
|---|---|---|---|---|---|
| | | $\mu_\circ$ | | | |
| $V_\circ(3)$ | $B_\circ(3)$ | Reg.2a | Reg.3a | Reg.2c | Reg.3c |
| 20.42 | 20.15 | $27.43 \pm 0.60$ | $27.82 \pm 0.91$ | $27.70 \pm 0.93$ | $28.24 \pm 1.18$ |

Pierce et al.'s use of the method gives $\mu_\circ = 27.0 \pm 0.2$. These distance estimates can be compared with the estimate obtained by Sandage et al. [16] using Hubble Space Telescope observations of Cepheids in IC 4182, $\mu_{AV} = 28.47 \pm 0.08$. After correction for foreground extinction using $A_{v\,I(100)} = 0.07 \pm 0.07$ $\mu_\circ = 28.40 \pm 0.11$. If this distance is combined with the $K$ band data of Pierce et al. then its position – shown in the $K$ band calibration of Figure 1 as the open square – is only $2.3\sigma$ from the regression line. In other words the Cepheid result is consistent with the brightest stars result because the uncertainties associated with the brightest stars method are so large.

## 6  Conclusions

1. Linear relationships exist for all calibration methods in the $V$, $K$ and $B$ bands.

2. All the relationships examined have a non-zero slope.

3. All the calibration methods give large regression errors which lead to $1\sigma$ distance modulus uncertainties in excess of 0.5 mag.

4. If the method is to be used then strict care must be taken to ensure that it is used properly: data should only be employed in bands for which a calibration exists and for appropriate classes of stars.

5. The method is unable to distinguish between the long and short distance scales and therefore should not be employed in this way.

**Acknowledgements.** Richard Rozanski acknowledges a grant from the SERC UK.